\documentclass[%
prl,twocolumn,
superscriptaddress,
 amsmath,amssymb,
 aps,
]{revtex4-2}

\usepackage{amsmath,amssymb,amsthm,mathrsfs,amsfonts,dsfont,amstext}
\usepackage{physics}
\usepackage{graphicx}
\usepackage{dcolumn}
\usepackage{bm}
\usepackage{booktabs} 
\usepackage{xcolor,ulem}
\usepackage{hyperref}
\hypersetup{
    colorlinks=true,
    linkcolor=blue,
    citecolor=blue,
    }

\newcommand{\yso}[0]{Y$_2$SiO$_5$}

\newcommand{\ybyso}[0]{$^{171}$Yb$^{3+}$:Y$_2$SiO$_5$}

\newcommand{\ybiso}[0]{$^{171}$Yb$^{3+}$}

\newcommand{\ybtransition}{$^2$F$_{7/2} \leftrightarrow ^2$F$_{5/2}$}

\begin{document}

\raggedbottom	

\title{Entanglement distribution and quantum storage of more than 8000 modes over a metropolitan network}

\author{A. Gelmini Rodriguez}
\affiliation{University of Geneva, Department of Applied Physics, Rue de l'Ecole-De-Médecine 20, 1205 Genève, Switzerland}

\author{L. Nicolas}
\affiliation{Department of Applied Physics, Aalto University, 02150 Espoo, Finland}

\author{T. Sanchez Mejia}
\affiliation{University of Geneva, Department of Applied Physics, Rue de l'Ecole-De-Médecine 20, 1205 Genève, Switzerland}

\author{P. Sekatski}
\affiliation{University of Geneva, Department of Applied Physics, Rue de l'Ecole-De-Médecine 20, 1205 Genève, Switzerland}

\author{N. Brunner}
\affiliation{University of Geneva, Department of Applied Physics, Rue de l'Ecole-De-Médecine 20, 1205 Genève, Switzerland}

\author{T. Taher}
\affiliation{University of Geneva, Department of Applied Physics, Rue de l'Ecole-De-Médecine 20, 1205 Genève, Switzerland}

\author{R. Thew}
\affiliation{University of Geneva, Department of Applied Physics, Rue de l'Ecole-De-Médecine 20, 1205 Genève, Switzerland}

\author{P. Goldner}
\affiliation{Chimie ParisTech, PSL University, CNRS, Institut de Recherche de Chimie Paris, Paris, France}

\author{M. Afzelius}
\email{mikael.afzelius@unige.ch}
\affiliation{University of Geneva, Department of Applied Physics, Rue de l'Ecole-De-Médecine 20, 1205 Genève, Switzerland}

\date{\today}

\begin{abstract}
Entanglement generation between telecommunication photons and matter is central to fibre-based quantum repeaters. Achieving practical communication rates requires multiplexing, which multimode quantum memories can provide. Rare-earth-ion ensembles offer large temporal multimode storage by exploiting the numerous spectral channels within their absorption spectrum. Here, we report on a quantum repeater node comprised of a \ybyso{} multimode quantum memory, featuring a 250 MHz bandwidth and a $76.6~\mu\mathrm{s}$ lifetime, and a bandwidth-matched entangled photon-pair source. We introduce and validate a quantitative measure of the effective temporal mode capacity using a Schmidt decomposition. With this platform, we demonstrate entanglement between a telecom photon propagating through a 25.3~km fiber spool and a 979 nm photon stored for $125~\mu\mathrm{s}$ across 16340 temporal modes. Finally, we report a field deployment distributing entanglement over 5.66~km through the Geneva metropolitan fibre network while storing 8235 modes for $63~\mu\mathrm{s}$.
\end{abstract}

\maketitle


\section{\label{sec:Intro}INTRODUCTION}

Future quantum networks hold the promise of quantum-enhanced technological capabilities in communication, computation, and sensing \cite{Wehner2018}. To achieve this vision, it is essential to distribute and store entanglement between remote quantum nodes. Fiber-based entanglement distribution can be achieved by a quantum repeater \cite{Briegel1998,Duan2001, Sangouard2011}, which leverages entanglement between quantum memories and telecom photons to connect remote locations. In this context, we have seen rapid experimental progress during this decade, across a range of different physical platforms, showing entanglement between telecom photons and matter \cite{Leent2020,LagoRivera2023,Rakonjac2023,Kucera2024,Liu2024,Krutyanskiy2024,Zhou2024,Bersin2024,Cui2025} and telecom-heralded matter-matter entanglement \cite{lago2021,Leent2022,Liu2024a,Knaut2024,Stolk2024,Haenni2025,zhu2026}, with several of these recent experiments carried out in standard metropolitan fiber networks \cite{Rakonjac2023,Liu2024a,Liu2024,Bersin2024,Kucera2024,Knaut2024,Stolk2024,zhu2026}.

One key challenge with quantum repeaters is the rate limitation imposed by the photon propagation time, about $5~\mu \mathrm{s/km}$ in silica optical fibers, when the quantum memory can only store a single excitation (or mode) \cite{Sangouard2011}. Quantum repeaters based on multiplexed quantum memories can achieve significantly higher rates by making many entanglement-generation attempts per round-trip time \cite{Collins2007, Simon2007}. With ensemble-based quantum memories, multiplexing can be achieved by storing photonic states into many time modes \cite{Businger2022, Rakonjac2023, Liu2024}, frequency modes \cite{Chakraborty2025} or spatial modes \cite{Zhang2024,Teller2025}, which is known as multimode storage. For some single quantum emitters, one can exploit multiqubit registers for multiplexing \cite{You2024, Krutyanskiy2024, Cui2025, Ruskuc2025, Canteri2025}. Whatever the physical platform and method, it is generally accepted that multiplexing is key to achieving practical entanglement generation rates in quantum repeaters.

\begin{figure*}
    \centering
    \includegraphics[width=1\linewidth]{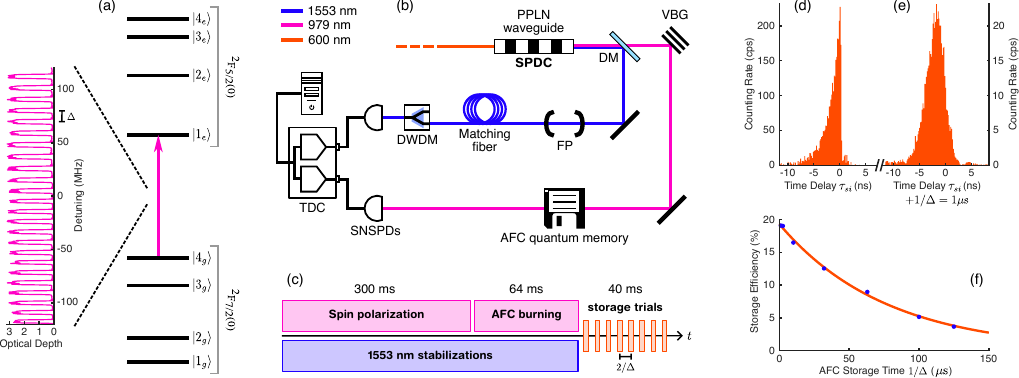}
    \caption{\textbf{Experimental setup}. (a) Hyperfine manifolds of the $^2$F$_{7/2}(0)$ ground and $^2$F$_{5/2}(0)$ excited states of \ybiso{} in crystallographic site II. The AFC memory was created on the $\ket{4_g} - \ket{1_e}$ transition, the inset shows a measured comb with periodicity $\Delta = 10~\mathrm{MHz}$. (b) Simplified schematics of the experimental setup showing the SPDC source (PPLN waveguide), the AFC quantum memory, the optical frequency filters, and the detection setup. The optical filters comprise the volume Bragg grating (VBG), 20 GHz bandwidth, Fabry-Pérot (FP) cavity, 90 MHZ bandwidth, and the dense wavelength-division mulitplexer (DWDM) at channel 31, $75$~GHz bandwidth. The detection system consists of superconducting nanowire single-photon detectors (SNSPDs) and a time-to-digital converter (TDC). DM = dichroic mirror. (c) The experimental measurement sequence, showing the memory preparation step (spin polarization and AFC burning), the frequency stabilization step, and the SPDC photon storage trials, see Methods for details. (d) Example of a detection histogram as a function of $\tau_{si}$, showing the correlation peak when the 979-nm signal photons pass a 250-MHz wide transmission hole in the absorption spectrum (reference measurement). (e) Example of a detection histogram recorded for an AFC periodicity of $\Delta = 1~\mathrm{MHz}$, showing 979-nm signal photons stored for $1~\mu\mathrm{s}$. (f) Measured AFC storage efficiency as a function of the storage time $1/\Delta$, based on measurements as shown in (e) and (f), see main text for the data analysis.}
    \label{fig:exp_setup}
\end{figure*}

In this work, we present a quantum repeater node consisting of a highly multimode \ybyso{} quantum memory and an entangled photon-pair source based on spontaneous parametric downconversion (SPDC). The memory is based on the atomic frequency comb (AFC) technique \cite{Afzelius2009a}, where the temporal multimode capacity depends on the product of the memory bandwidth and the optical storage time \cite{Ortu2022}, the latter ultimately being limited by the optical coherence time. The rare-earth-doped crystal \ybyso{} is particularly interesting in this context, owing to the hybridized electron-nuclear hyperfine states in \ybiso{} that possess long optical coherence times at cryogenic temperatures, i.e. they are natural "clock" states \cite{Ortu2018,Kindem2020,Nicolas2023}. The electronic spin component also creates large hyperfine splittings in the GHz range, enabling broadband memory operation. Exploiting these unique features, we recently developed a classical AFC memory in \ybyso{} with a bandwidth of 250 MHz and storage times reaching $125~\mu\mathrm{s}$, as characterized by narrowband laser pulses in Ref. \cite{SanchezMejia2025}. Here, we demonstrate broadband storage of quantum entanglement across a large number of temporal modes, through storage of a 979-nm photon entangled with a 1553-nm telecom photon. By employing the technique of Schmidt decomposition of the joint temporal amplitude (JTA) of the two-photon state, one can quantify the effective multimode capacity, which ranges from 327 to 16340 modes, corresponding to the different memory storage times, which were set equal to the photon propagation time for each fiber distance (ranging from 0.5 to 25.3 km).

\section{\label{sec:Results}RESULTS}
\subsubsection{Experimental setup}

The core parts of the experimental setup are the AFC quantum memory and the SPDC source of entangled photon pairs, see Fig. 1. The AFC is implemented with an yttrium orthosilicate crystal, \yso{}, doped with 2 ppm of \ybiso{} ions, which is cooled to about 3 K in a closed-cycle cryostat. The AFC memory is based on a comb-like high-contrast modulation of the absorption profile, with periodicity $\Delta$ in the frequency domain, to achieve storage for a pre-determined duration of $1/\Delta$. The AFC is created through optical pumping on the \ybtransition{} transition in \ybiso{}, see Fig. \ref{fig:exp_setup}(a), at the wavelength of 979~nm, using a frequency-agile and broadband optical setup described in detail in Ref. \cite{SanchezMejia2025}. 

The SPDC source is based on a periodically poled lithium niobate (PPLN) waveguide, see Fig. \ref{fig:exp_setup}(b). The crystal is pumped with a monochromatic continuous-wave laser at 600 nm, and it is phase-matched to generate photon pairs at 979~nm (signal mode) and in the telecom C-band at 1553~nm (idler mode). To match the bandwidth of the memory, the SPDC emission in both modes is frequency filtered. The telecom mode is filtered by a narrowband Fabry-Perot cavity with a full-width at half-maximum (FWHM) linewidth of 90~MHz, followed by a dense wavelength division multiplexer (75 GHz bandwidth), which selects a single longitudinal mode of the cavity. The 979-nm signal mode is coarsely filtered (20 GHz bandwidth) by a volume Bragg grating before the memory and the final filtering is done by the 250 MHz AFC memory. Conceptually, the detection of the telecom photon, after propagation through the fiber, heralds the storage of a bandwidth-matched 979-nm photon in the memory, which is the key feature allowing generation of remote entanglement in a quantum repeater scheme, see eg. Refs \cite{Simon2007,Sangouard2011}.

The experimental sequence consists of two phases: a preparation phase of 364 ms and a measurement phase of 40~ms, see Fig. \ref{fig:exp_setup}(c). During the preparation, the \ybiso{} ions are spin-polarized into a single hyperfine state to increase the optical depth (300~ms), and the AFC is created through frequency-selective optical pumping (64~ms), as described in detail in Ref. \cite{SanchezMejia2025}. During the measurement phase, the SPDC source is pumped for a duration $T_\text{pump}$, which is set equal to the storage time, $T_\text{pump} = 1/\Delta$, to maximize the number of stored temporal modes. After the SPDC pump is turned off, the memory emits (read-out) all the stored modes for another period of $1/\Delta$, such that the total duration of the storage (pump on) and memory read-out (pump off) takes $2/\Delta$. This sequence is then repeated such that the total measurement phase lasts 40~ms.

The idler and signal photons are detected by efficient superconducting nanowire single photon detectors (SNSPDs) and all detections were logged by a time-tagger for post-processing. From the time-tagged detections, one can compute the cross-correlation histograms, eg. Figs \ref{fig:exp_setup}(d)-(e), or the normalized second-order cross-correlation function, $g_{si}^{(2)}(\tau_{si})$. To ensure a high cross-correlation, an active frequency stabilization scheme ensures that the filtered idler and signal photons conserve energy with respect to the pump, see Methods.

\begin{figure*}
    \centering
    \includegraphics[width=1\linewidth]{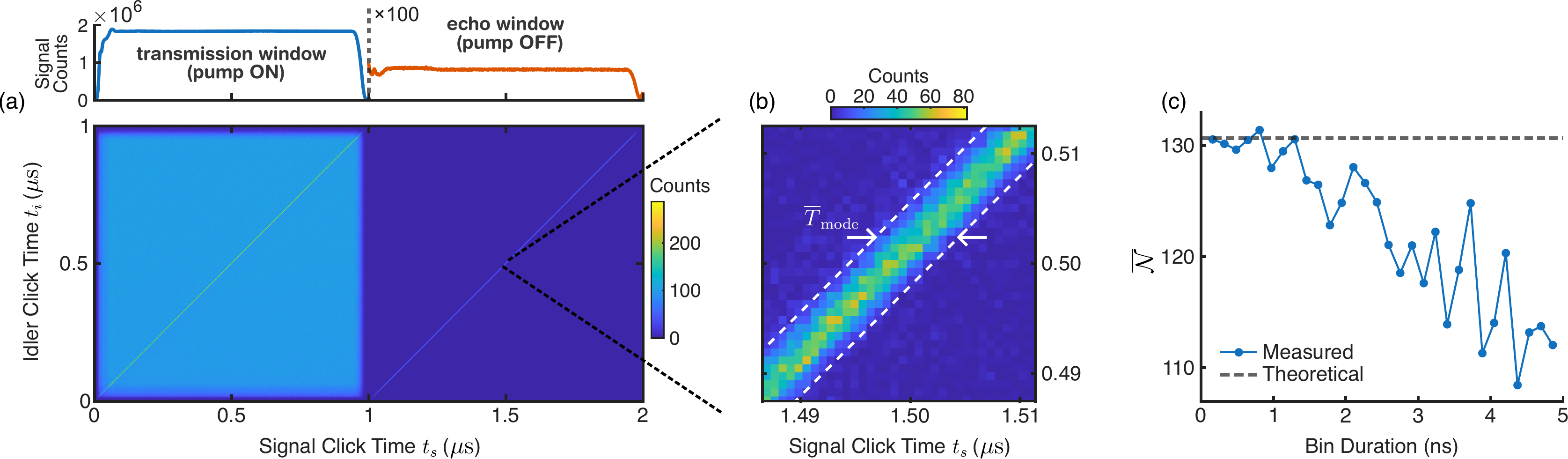}
    \caption{\textbf{Multimode analysis.} (a) The measured joint detection matrix as a function of the idler $t_i$ and signal $t_s$ detection times, with respect to the beginning of the SPDC pump pulse. The background due to accidental coincidences is significantly higher in the transmission window, due to 979-nm signal photons not absorbed by the \ybyso{} crystal. The signal-to-noise ratio is significantly higher in the echo window, owing to the frequency-temporal filtering by the AFC quantum memory \cite{Rielander2014}. (Top inset) Total signal counts as a function of $t_s$ showing the transmission and echo windows. (b) Zoom on the echo correlation line, at a binning of 730 ps. (c) Numerical estimation of the lower bound of the Schmidt mode number $\overline {\mathcal{N}}$, based on the measured cross-correlation function $g_{si}^{(2)}(\tau_{si})$, as a function of the data binning size used when computing $g_{si}^{(2)}(\tau_{si})$. The dashed line represents the expected theoretical value of $\overline {\mathcal{N}}$, see text for details}.
    \label{fig:mm_analysis}
\end{figure*}

\subsubsection{Characterization of the broadband quantum memory}

The memory efficiency $\eta_{\mathrm{AFC}}$ was measured for varying AFC storage times $1/\Delta$, ranging from $1~\mu\mathrm{s}$ to $125~\mu\mathrm{s}$, shown in Fig. \ref{fig:exp_setup}(f). The efficiency was computed from the cross-correlation histogram taken with the AFC, and a reference histogram measured with a $250$-MHz wide transmission hole burnt into the same optical transition. At short storage times, the measured efficiency $\eta_0 = (19 \pm 1)\%$ is in good agreement with the expected efficiency for an optimal square-shaped AFC, essentially only limited by the finite optical depth of the sample. It is similar to the efficiency obtained with narrowband classical pulses \cite{SanchezMejia2025}, showing that the SPDC source is well bandwidth-matched to the quantum memory and an accurate active frequency stabilization scheme for the 979-nm SPDC photon.

The memory efficiency decays exponentially with the storage duration, $\eta_\mathrm{AFC} = \eta_0 \exp(-1/(\Delta T_M))$, with a characteristic memory time $T_M = (76.6 \pm 11.0)~\mu\mathrm{s}$. This corresponds to an effective AFC coherence lifetime \cite{Jobez2016} of $T_2^{\mathrm{AFC}}= 4T_M = (307 \pm 44)~\mu\mathrm{s}$, the longest achieved in the quantum regime in any AFC memory, to our knowledge. Yet, it is shorter than the 1 ms optical coherence time \cite{Nicolas2023}, showing the potential for further progress.

\subsubsection{Multimode analysis through Schmidt decomposition}

When an SPDC source is pumped by a long laser pulse with respect to the photon pair coherence time, it will produce photon pairs in many temporal modes, of the order of the ratio of the pump pulse duration to the coherence time. The multimode character of the storage can be visualized by computing a joint detection matrix, i.e. the joint detection probability as a function of the relative delays of the signal and idler detection times with respect to the start of the pump pulse, as shown in Fig. \ref{fig:mm_analysis}(a). This particular experiment was performed without any fiber spools in the idler mode and with a short memory storage time of $1~\mu s$, with the goal of gathering enough statistics.

The joint detection matrix shown in Fig. \ref{fig:mm_analysis}(a) displays the expected strong correlation, seen as two clear diagonal correlation lines. The correlation line detected between $t=0$ and $t=1/\Delta$, i.e., when the pump is on, is due to idler and signal photons being detected simultaneously, where the 979~nm signal photon is transmitted through the memory due to its non-unit absorption efficiency. The second correlation line between $t=1/\Delta$ and $t=2/\Delta$, i.e., when the pump is off, is due to the joint detection of an idler photon and a stored 979 signal photon, detected during the QM read-out phase after a delay of $1/\Delta = 1~\mu \mathrm{s}$. The high multimode capacity can already be qualitatively observed from the extremely narrow correlation line.

To gain more insight into the multimode capacity, one can perform a Schmidt decomposition of a two-photon state 

\begin{equation}
       \ket{\Psi} \propto \int \dd t_s \,\dd t_i\, \Phi(t_s,t_i) \, a^\dag(t_s) \, b^\dag(t_i)\ket{0},
       \label{eq:state}
\end{equation}

\noindent with $a^\dag$ and $b^\dag$ denoting the signal and idler spatial modes, respectively. Crucially, $\Phi(t_s,t_i)$ is the complex JTA, and $|\Phi(t_s,t_i)|^2$ is the joint temporal intensity (JTI) \cite{Kuzucu2008,MacLean2018, Borghi2024, Yu2024}. Note that the JTI can be extracted from the measured joint detection matrix, by subtracting the background of accidental coincidences (see discussion below).

In the case of a long pump pulse, with a slowly varying temporal envelope $|\alpha(t)|$, the JTA can be factorized $\Phi(t_s,t_i) = \alpha(t_i) \phi(t_s-t_i) = \alpha(t_i) \phi(\tau_{si})$ with $\tau_{si} = t_s-t_i$. In this case, the JTA only depends on the relative function $\phi(\tau_{si})$, where the measured normalized cross-correlation function $g_{si}^{(2)}(\tau_{si})$ is proportional to $|\phi(\tau_{si})|^2$ for the state in Eq. \eqref{eq:state}.

A Schmidt decomposition of the JTA of the pure state in Eq. \eqref{eq:state} allows calculating the effective number of modes $\mathcal{N}$ (the Schmidt mode number) through the purity of the reduced state \cite{Law2000,law2004analysis}, as shown in the Supplementary Information (SI). Note that we are not seeking to quantify high-dimensional entanglement, which also requires considering the phase coherence over the modes. For the special case of a long pump pulse with rectangular amplitude envelope, the Schmidt mode number is (see Eq. (23) in the SI)

\begin{equation}\label{eq:Schmidt_N}
\mathcal{N} =T_\text{pump} \frac{\left( \int  \dd{\tau_{si}}\abs{\phi(\tau_{si})}^2 \right)^2}{\int \dd{\tau_{si}}\abs{\int \dd s   \phi(s) \phi^*(s-\tau_{si}) }^2}.
\end{equation}

\noindent Note that the corresponding formula for the joint spectral amplitude defined in the frequency domain is given in the SI. 

Calculating $\mathcal{N}$ requires phase information of the JTA of the corresponding state. But, from an experimentally measured JTI, there is no phase information and one must then assume JTA to have a constant phase. However, it can be shown that calculating the Schmidt mode number from the real quantity $|\phi(\tau_{si})|$ provides a lower bound

\begin{equation}\label{eq:Schmidt_N_lb}
 \overline {\mathcal{N}} =T_\text{pump} \frac{\left(\int  \dd{\tau_{si}}\abs{\phi(\tau_{si})}^2 \right)^2}{\int \dd{\tau_{si}}\abs{\int \dd s   |\phi(s)| \times |\phi^*(s-\tau_{si})| }^2} \leq \mathcal{N}.
\end{equation}

\noindent Note that for a JTA with factorizable phases, these two expressions give the same result $\overline {\mathcal{N}} = \mathcal{N}$. One can also naturally define an effective temporal mode duration $T_\text{mode}$ ($\overline{T}_\text{mode}$), by introducing $\mathcal{N} = T_\text{pump} / T_\text{mode}$ ($\overline {\mathcal{N}} = T_\text{pump} / \overline{T}_\text{mode}$).

To summarize, one can compute the theoretical Schmidt mode number from the JTA of the two-photon state, retaining full phase information, through Eq. \eqref{eq:Schmidt_N}. Analytical expressions of $\mathcal{N}$ for two-sided Lorentzian or Gaussian spectral functions are given in the SI. The former is appropriate for cavity-enhanced SPDC sources, eg. Refs \cite{lago2021,Haenni2025}. Note that for both these cases the phases are factorizable, hence $\overline {\mathcal{N}} = \mathcal{N}$. For our experimental set-up with a rectangular-shaped signal filter (the AFC memory) and a Lorentzian idler filter (the FP cavity), the phases are not factorizable. The analytical formula for $\mathcal{N}$ is given in the SI, while the lower bound $\overline {\mathcal{N}}$ can be calculated numerically from Eq. \eqref{eq:Schmidt_N_lb}.

Finally, we can estimate the Schmidt mode number directly from the measured cross-correlation function $g_{si}^{(2)}(\tau_{si})$ through the lower bound, Eq. \eqref{eq:Schmidt_N_lb}. However, the measured state is a two-mode squeezed vacuum state generated through the SPDC process, which has higher order multipair contributions not taken into account in Eq. \eqref{eq:state}. These generate accidental coincidences that appear as a background with mean value of 1 in the measured $g_{si}^{(2)}$ function. Therefore, we subtract the background and approximate $|\phi(\tau_{si})| \propto   \sqrt{g_{si}^{(2)}(\tau_{si})-1}$. The results are shown in Fig. \ref{fig:mm_analysis}(c), as a function of the binning size of the data. The experimental Schmidt mode number approaches the numerically calculated theoretical lower bound $\overline {\mathcal{N}}$ for small binnings and stays below the theoretical Schmidt mode number $\mathcal{N}$ for this particular state, as expected (see SI for more details).

The theoretical Schmidt mode numbers given later in the article were all based on the more conservative lower bound $\overline {\mathcal{N}}$ for our particular experiment. Moreover, the integration window used for all the entanglement experiments was based on the corresponding effective mode duration $\overline{T}_\text{mode} = 7.65~\mathrm{ns}$, which contains ${\sim}94\%$ of the photons in the $g_{si}^{(2)}$ correlation peak.

\subsubsection{Entanglement distribution over lab fiber spools}

\begin{figure*}
    \centering
    \includegraphics[width=1\linewidth]{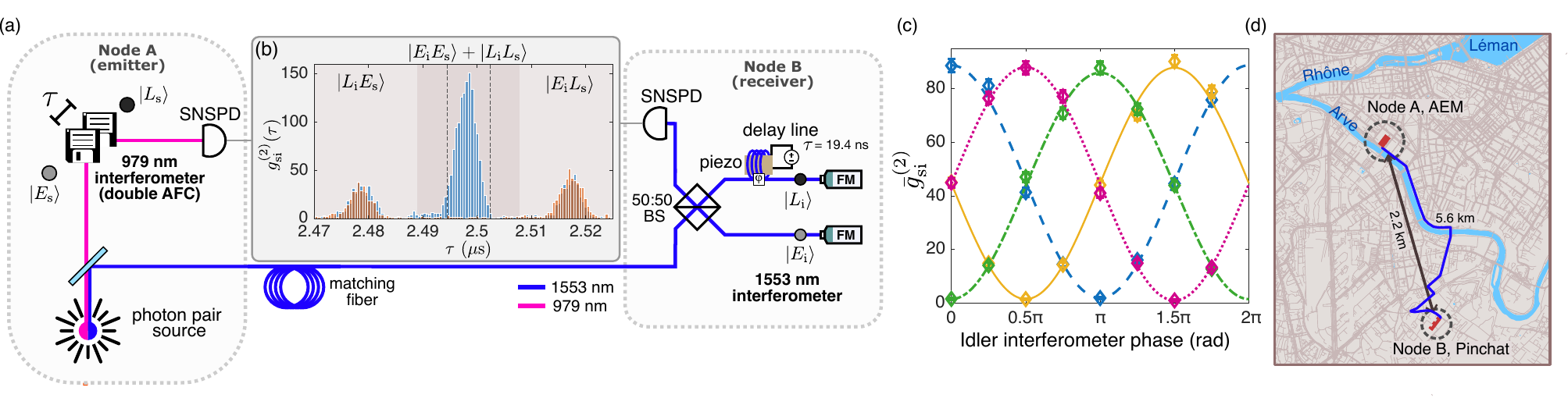}
    \caption{\textbf{Entanglement distribution.} (a) Franson-type interferometer setup for measuring energy-time entanglement, employing a fibered Michelson interferometer at 1553~nm and a double-AFC scheme at 979~nm. FM = Faraday mirror. BS = beam splitter. (b) Example of detection histogram for $1/\Delta = 2.5~\mu\mathrm{s}$, showing the typical three correlation peaks of the Franson setup, where only detections in the middle peak reveal entanglement. The blue and red histograms show cases of constructive and destructive interferences. Dashed lines indicate the detection window $\overline{T}_\text{mode}$. (c) The averaged second-order cross-correlation value $\bar{g}_{si}^{(2)}$ as a function of the idler phase $\phi_i$, for fixed signal phases $\phi_s = 0$ (blue), $\pi/2$ (yellow), $\pi$ (green) and $3\pi/2$ (red). (d) In a field test, Node B was moved to another university building 2.15~km away, connected to Node A through the Geneva metropolitan network using a standard, dark optical fiber of length 5.66~km.}
    \label{fig:Bell_test}
\end{figure*}

The photon pairs generated through SPDC with a coherent continuous pump are energy-time entangled \cite{Xavier2025}, which can be measured using a Franson-type interferometer setup using two unbalanced interferometers \cite{Franson1989, Tittel1999, Xavier2025}, see Fig. \ref{fig:Bell_test}(a). The coincidence detection histogram after the interferometers display three peaks, see Fig. \ref{fig:Bell_test}(b), where a detection in the middle peak post-selects an entangled state \cite{Tittel1999}

\begin{equation}
    \ket{\Phi} = \frac{1}{\sqrt{2}}(\ket{E_s E_i}+\ket{L_s L_i}),
\end{equation}

\noindent where $\ket{E_{s, i}}$ and $\ket{L_{s, i}}$ denote the post-selected early and late time bins, for the idler and signal, respectively. The time-bin separation is given by the identical delay in the unbalanced interferometers, here set to $19.4$~ns. The telecom idler photon is analyzed with a phase-stabilized fiber-based Michelson interferometer, and the signal 979~nm photon is analyzed using the double-AFC scheme \cite{Riedmatten2008}. The double-AFC lowers the memory efficiency by a factor of 1/4 for the entanglement measurements, however, it has the advantage of not requiring an additional phase stabilization scheme, as the phase is given by the relative detunings of the two AFCs \cite{Afzelius2009a}. Both analyzers thus have controllable phases $\phi_i$ and $\phi_s$, respectively.

\begin{table*}[]
    \resizebox{1.25\columnwidth}{!}{
    \begin{tabular}{@{}c|c|c|c|c|c|c|c|c@{}}
        \toprule
        $1/\Delta$ & $L_f$ & Fiber loss & $S$ & $\bar{g}_{si}^{(2)}$    &  $\overline {\mathcal{N}}$ & $R_i$ (cps) & $R_s$ (cps) & $R_c$ (cps) \\ \midrule   
        2.5 $\mu$s & 0.5 km   & 1.22 dB & 2.72(3) & 83  & 327   & 50000 & 14000 & 450 \\
        10 $\mu$s  & 2.005 km & 0.77 dB & 2.73(4) & 85  & 1307  & 44000 & 13000 & 360 \\
        32 $\mu$s  & 6.521 km & 1.86 dB & 2.73(3) & 99  & 4183  & 54000 & 9000  & 368 \\
        63 $\mu$s  & 12.9 km  & 2.88 dB & 2.68(4) & 94  & 8235  & 44000 & 6400  & 199 \\ 
        125 $\mu$s & 25.3 km  & 5.00 dB & 2.70(4) & 99  & 16340 & 24000 & 2500  & 49  \\ \midrule   
        32 $\mu$s  & 5.66 km  & 3.83 dB & 2.65(4) & --  & 4183  & --    & --    & --  \\
        63 $\mu$s  & 5.66 km  & 3.83 dB & 2.67(4) & 116 & 8235  & 19000 & 5200  & 87  \\ \bottomrule 
    \end{tabular}
    }
    \caption{\textbf{Summary of results.}  Results for the entanglement distribution experiments, including experiments with fiber spools (top) and the field experiment (bottom). Shown are the AFC memory time $1/\Delta$, the fiber length $L_f$ and loss, the measured CHSH $S$ parameter. $\bar{g}_{si}^{(2)}$ is the average value of $g_{si}^{(2)}(\tau_{si})$ over the detection window. $\overline {\mathcal{N}}$ is the lower bound of the number of stored Schmidt modes. The idler ($R_i$), signal ($R_s$) and coincidence ($R_c$) rates were measured without the CHSH analysers (values for the 32 $\mu$s field experiment were not available).}
    \label{Tab:Bell_test_fibers}
\end{table*}

Coincidence histograms were computed as a function of the relative detection time of the idler and signal photons, only considering $979$~nm photons detected during the QM read-out phase. The coincidence rate $R$ depends on the measurement phases as $R(\phi_i,\phi_s) = A(1+V \cos(\phi_i+\phi_s))$. Fig. 3(b) shows an example of the histogram at the optimized phases for constructive and destructive interference, for the memory storage time of $1/\Delta = 2.5~\mu\mathrm{s}$. A scan of the phase $\phi_i$ produces an interference fringe with a visibility of $V = (97 \pm 1)\%$, see Fig. 3(c), from the raw data without subtracting dark counts or accidental coincidences, which is a strong indication of entanglement preservation during the storage in the quantum memory.

To demonstrate storage of entanglement, one can perform a Clauser-Horne-Shimony-Holt (CHSH) Bell test~\cite{Clauser1969} on the post-selected coincidence data. Under the fair-sampling assumption the latter can reveal Bell nonlocality of the observed correlations~\cite{orsucci2020post}, which we use as a robust witness of the presence of entanglement. The coincidence rate $R(\phi_i,\phi_s)$ was measured for the standard CHSH settings $\phi_i = \pi/4$, $\phi_i = -\pi/4$, $\phi_s = 0$ and $\phi_s = \pi/2$, from which we calculated the CHSH $E(\phi_i,\phi_s)$ correlators and the $S$ parameter. For separable states, the CHSH inequality must satisfy $S\leq 2$, and values between $S = 2$ and $S = 2\sqrt{2}$ can only be obtained with entangled states. The CHSH settings above maximizes the S parameter. Note that both analysers only had a single output port, hence in addition, we had to measure $\phi_i + \pi$ and $\phi_s + \pi$ to access the second measurement outcome. For each measurement setting, the coincidence histogram peak was averaged over a detection window of $\overline{T}_\text{mode} = 7.65~\mathrm{ns}$. 

The CHSH test was performed with fiber spools of varying lengths, $L_f$, ranging from 0.5 to 25.3~km, placed in the telecom idler mode between the SPDC source and the analyzing interferometer. For each fiber spool, the quantum storage time $1/\Delta$ was set to slightly longer than the propagation time in the fiber spool. In Table \ref{Tab:Bell_test_fibers}, we summarize the results of the CHSH test as a function of fiber length. The $S$ parameter violates the CHSH inequality for all fiber lengths, proving the presence of entanglement after storage of the 979~nm photon and the propagation of the telecom photon for distances of up to 25~km.

The $S$ parameter obtained for $1/\Delta = 2.5~\mu\mathrm{s}$ and $L_f = 0.5~\mathrm{km}$ can be compared to the visibility curve recorded for the same settings. The $S$ parameter is expected to vary as $S = 2\sqrt{2} V$, resulting in an expected CHSH parameter of $S = 2.73 \pm 0.03$, for the measured visibility, cf. Fig. 3(c), in close agreement with the measured $S$ parameter, showing that there is no unexpected correlated noise in the measurements.

In Table \ref{Tab:Bell_test_fibers}, we also show the detection rates obtained for various fiber lengths and storage times, measured without the Michelson interferometer and using a single AFC with the storage efficiency shown in Fig. \ref{fig:exp_setup}(f). The telecom idler detection rate varies little for fiber losses below 3~dB, due to variations of the pair production rate and coupling losses across the different measurements, the only clear difference being the idler rate for the lossiest fiber of 25.3~km. The signal and coincidence rates clearly drop across the measurements, which is due to the exponential decrease in memory efficiency. We also report the average second-order cross-correlation function $\bar{g}_{si}^{(2)}$, averaged over the detection window of $\overline{T}_\text{mode}$. The cross-correlation value stays constant across all measurements, implying that it is limited by the photon-pair creation probability of the SPDC source. The visibility of the CHSH test is expected to follow $V = (\bar{g}_{si}^{(2)}-1)/(\bar{g}_{si}^{(2)}+1)$ and the CHSH $S$ parameter should follow $S=2\sqrt{2}V$. The measured cross-correlation values $\bar{g}_{si}^{(2)}$ imply an expected CHSH parameter between $S = 2.76$ and $2.77$, in good agreement with the measured $S$ parameters. The small difference can be attributed to the accuracy of the phase locking scheme of the Michelson interferometer.

As the storage time increases, the number of stored temporal modes increases correspondingly. In Table \ref{Tab:Bell_test_fibers}, we report the Schmidt mode number $\overline {\mathcal{N}}$ based on the mode size $\overline{T}_\text{mode} = 7.65~\mathrm{ns}$. The mode capacity reaches more than 16340 modes at $125~\mu\mathrm{s}$ storage time, an improvement of one to two orders of magnitude with respect to other works \cite{TangZhouWangEtAl2015,lago2021,Rakonjac2023,Businger2022,Wei2024,Liu2024,zhu2026}. Yet, there is a clear trade-off between coincidence detection rate and the storage time, due to the exponential decay of the memory efficiency. Still, the memory can store 1307 modes for a storage time of $10~\mu\mathrm{s}$, with a minimum penalty for the memory efficiency, which reaches 88\% of its maximum value as given by the optical depth. It should also be emphasized that the AFC coherence time is only about $1/3$ of the optical coherence time, implying that in the future these numbers can be improved by up to a factor of $3$. Only in a system having both large bandwidth and long optical coherence can one in principle achieve high efficiency and high multimode storage, a clear advantage of \ybyso{} with respect to other rare-earth doped crystals.

\subsubsection{Entanglement distribution over the Geneva metropolitan network}

In a more realistic quantum repeater setup, the telecom SNSPD detector system, the DWDM filter and the Michelson interferometer and associated control electronics were moved to another university building, site B, while the quantum memory and the SPDC source stayed at site A, see Fig. \ref{fig:Bell_test}(d). The sites were physically separated by 2.15~km and connected via a standard dual-strand single-mode optical fiber measuring 5.66~km within the Geneva metropolitan network. One fiber strand was used as the dark quantum channel and the other fiber strand was used as a classical synchronization channel (service channel). A clock signal was sent from site A to B through the service channel, using a pulsed telecom laser, allowing synchronization of both the time tagger and the interferometer phase stabilization circuit at site B.

A CHSH Bell test was performed in the metropolitan setting, for two different storage times of $1/\Delta = 32~\mu \mathrm{s}$ and $63~\mu \mathrm{s}$, corresponding to slightly more than a single-trip and round-trip time of the 5.66-km long fiber link, respectively. The measured CHSH parameters were $S = 2.65 \pm 0.04$ and $S = 2.67 \pm 0.04$, respectively, clearly demonstrating the generation of light-matter entanglement over a 2.15~km straight-line distance. The two memory storage times correspond to $\overline {\mathcal{N}} = 4183$ and 8235 stored modes.

\section{\label{sec:Summary}Discussion}

The experiments reported in this work constitute half of an elementary quantum repeater link. In particular, the telecom detection rates we achieve in this work, in the kcps regime, are representative of the entanglement heralding rates expected (for the elementary link) in repeater schemes based on single-photon entanglement \cite{Sangouard2011}, so called DLCZ-type repeaters, made possible by the broadband and multimode \ybyso{} quantum memory.

A key aspect of this work is the highly multimode storage, which promises a great speed-up of a quantum repeater. In this work, we introduced the Schmidt mode number as an effective mode number, but it is an open question if this number represents the expected speed-up with respect to a single-mode repeater. In the SI, we show that for a simplified model based on the heralding (conditioning) on the idler detection-time, and perfectly filtering (projecting on) the conditional modes of the signal photon after the memory, the speed-up factor is exactly the Schmidt mode number. However, in practice filtering the conditional modes is infeasible, and post-memory photons are filtered with a rectangular detection time-window resulting in some excess noise due to uncorrelated photons. In addition, the entanglement distribution rate depends on many parameters, in particular the SPDC pump power, the aforementioned detection time-window, the repeater swap depth (nesting level). A recent work~\cite{Hellebek2025} presented a detailed study of the expected rate of DLCZ-type repeaters, taking into account these aspects, both for pulsed and continuously pumped SPDC sources. A highly interesting extension of such calculations would be to look at the speed-up of the multimode case (continuous pump) with respect to the ideal single-mode case (infinitely short pump pulse) and to compare it to the effective Schmidt mode number introduced here. We hypothesize that the Schmidt mode number of the two-photon state would allow a simple quantitative benchmarking of different quantum repeater experiments in terms of multimode capacity and its potential speed-up.

To achieve a scalable quantum repeater based on \ybyso{}, several key points need to be addressed. The memory storage duration should at least reach the millisecond regime \cite{Wu2020}, and it should be tunable during storage (on-demand read out). This is possible through the spin-wave AFC scheme \cite{Afzelius2009a}, which has been demonstrated in \ybyso{} for up to 100 MHz bandwidth in the classical regime, for storage times up to a few milliseconds \cite{Businger2020, Businger2023}. The main challenge is to achieve spin-wave storage at the single-photon regime, by reducing noise generated by the intense control fields. The efficiency should also be improved, by exploiting low-finesse cavity in the impedance-matching regime, eg. Refs. \cite{Duranti2024,Davidson2020}. Attaining these objectives would allow a scalable quantum repeater based on \ybyso{}.

\appendix
\section{METHODS}

\textbf{The \ybyso{} quantum memory.} The \yso{} crystal was doped with 2 ppm of $^{171}$Yb$^{3+}$ ions, with an isotopic purity of 95$\%$. The crystal was cut along the D$_1$, D$_2$, b axes \cite{Li1992}, with dimensions 15.8, 3.9 and 3.1 mm. We work with ions occupying crystallographic site II, at the transition wavelength of 978.87 nm for the $^2$F$_{7/2}(0)$ - $^2$F$_{5/2}(0)$ transition. The photons passed the crystal four times to increase the optical depth, which was $d = 3.2$ after the spin polarization step in the memory preparation sequence. The memory setup and the memory preparation sequence is described in detail in Ref. \cite{SanchezMejia2025}.

\textbf{The entangled SPDC source} The source is based on the same approach as in \cite{Businger2022}. A $600.35$~nm beam weakly pumps a $\chi^{(2)}$ PPLN waveguide, generating photon pairs via SPDC. A first dichroic mirror filters out the pump, and a second separates the $978.87$~nm \textit{signal} mode and the telecom-band $1552.52$~nm \textit{idler} mode. The idler mode is filtered by a FP cavity with $90.3$~MHz FWHM (free spectral range $55$~GHz) and a channel 31 DWDM with $75$~GHz FWHM. The signal mode is filtered by a $20$~GHz VBG, the $250$~MHz AFC envelope, and a $5$~nm bandpass filter. The $600.35$~nm pump is generated via sum-frequency generation (SFG) in another PPLN waveguide using $978.87$~nm and $1552.52$~nm pumps that are frequency-locked to the center of the AFC and the FP, respectively. This locking scheme ensures that the filtered idler and signal modes are frequency-correlated. The FP cavity and both PPLN waveguides are temperature-stabilized using Peltier elements.

\textbf{Experimental sequence} The sequence is synchronized to the compressor's $700$~ms cycle. $300$~ms are dedicated to the spin polarization step, then $64$~ms to the AFC-burning. Storage trials then last for $40$~ms, with the rest of the cycle spent idling. During the idle time and the $364$~ms preparation duration, the $1552.52$~nm pump laser is frequency-locked to the FP by a field-programmable gate array acting as a proportional–integral–derivative regulator, using a Pound–Drever–Hall error signal and an optical switch to bring the pump from the SFG waveguide to the FP. When using the telecom Franson interferometer, only the idling-time is spent locking to the FP, with the $364$~ms spent stabilizing the phase of the interferometer using a side-of-fringe error signal generated by injecting an attenuated fraction of the $1552.52$~nm laser through it and counting SNSPD clicks using a microcontroller, which acts as regulator to feed back voltage onto the piezoelectric tube around which the long-path fiber is coiled. The lock setpoint is re-calibrated every $15$ minute by scanning the interferometer phase and finding the number of clicks corresponding to the midpoint of the interference fringe. Every minute, the $600.35$~nm pump power is stabilized to $300~\mu\mathrm{W}$ by measuring its power transmitted through the SPDC waveguide and feeding back onto the amplitude of an acousto-optic modulator in the $978.87$~nm laser beam path. Throughout the sequence, multiple mechanical shutters are used to protect the SNSPDs from bright light.

\textbf{Photon detectors} All single-photon detection events were recorded using superconducting nanowire single-photon detectors (SNSPDs) housed in separate closed-cycle cryostats at the two nodes. The detector at Node A operated at 750 mK, with $\sim$70\% detection efficiency, an $\sim$30 cps dark count rate and a jitter of 37~ps at 979 nm. The Node B detector operated at 900 mK, with a detection efficiency of $\sim$85\%, a dark count rate of $\sim$60 cps and a jitter of 41~ps at 1553 nm.

\section{ACKNOWLEDGEMENTS}

We acknowledge funding from the Swiss State Secretariat for Education, Research and Innovation (SERI) under Contract Numbers UeM029-7 and UeM019-3. P. G. further acknowledges funding from the France 2030 program (ANR-22-PETQ-0010, project QMemo). We also thank the OSCIN office at the Canton de Gen\`{e}ve for giving us access to the metropolitan fiber link.

\bibliographystyle{apsrev4-2} 
%


\end{document}